\documentclass[preprint,aps]{revtex4-1}
\usepackage{graphicx, bm, amssymb}
\usepackage{hyperref, cleveref}
\usepackage{setspace, float}
\usepackage[fleqn]{amsmath}
\usepackage{epstopdf}

\begin{document}
\def\be{\begin{equation}}
\def\ee{\end{equation}}
\def\ba{\begin{array}}
\def\ea{\end{array}}
\def\bea{\begin{eqnarray}}
\def\eea{\end{eqnarray}}
\def\bc{\begin{center}}
\def\ec{\end{center}}

\title{Study of rare semileptonic $B_c^+ \to D^+ \nu \bar{\nu}$ decay in the light-cone quark model}
\author{Nisha Dhiman and Harleen Dahiya}
\affiliation{Department of Physics,\\ Dr. B.R. Ambedkar National
Institute of Technology,\\ Jalandhar, 144011, India}

\begin{abstract}
We study the exclusive semileptonic rare $B_c^+ \to D^+ \nu \bar{\nu}$ decay in the framework of light-cone quark model. The transition form factors $f_{+}(q^2)$ and $f_{T}(q^2)$ are evaluated in the time-like region using the analytic continuation method in $q^{+} = 0$ frame. The analytic solutions of these form factors are compared with the results obtained from the double pole parametric form. The branching ratio for $B_c^+ \to D^+ \nu \bar{\nu}$ decay is calculated and compared with the other theoretical model predictions. The predicted results in this model can be tested at the LHCb experiments in near future which will help in testing the unitarity of CKM quark mixing matrix, thus, providing an insight into the phenomenon of CP violation.
\end{abstract}
\maketitle

\section{Introduction}
In the past few years, great progress has been made in understanding the semileptonic decays in the $B$ sector as these are among the cleanest probes of the flavor sector of the Standard Model (SM) which not only provide valuable information to explore the SM but are also powerful means for probing different new physics (NP) scenarios beyond the SM (BSM) \cite{dingfelder, choi1, jaus1}. Due to the Glashow-Iliopoulos-Maiani (GIM) mechanism \cite{glashow}, flavor changing neutral current (FCNC) induced semileptonic $B$ decays are rare in the SM because these decays are forbidden at tree level and can proceed at the lowest order only via electroweak penguin and box diagrams \cite{ali, buch}. Therefore, these decay processes provide sensitive probes to look into physics BSM \cite{blake}. They also play a significant role in providing a new framework to study the mixing between different generations of quarks by extracting the most accurate values of Cabibbo-Kobayashi-Maskawa (CKM) matrix elements which help us to test the charge-parity (CP) violation in the SM and to dig out the status of NP \cite{kim, aliev}.
\par The theoretical analysis of CP violating effects in rare semileptonic $B$ decays requires knowledge of the transition form factors that are model dependent quantities and are scalar functions of the square of momentum transfer \cite{wick}. These form factors also interrelate to the decay rates and branching ratios of all the observed decay modes of $B$ mesons and their calculation requires a non-perturbative treatment. Various theoretical approaches, such as relativistic constituent quark model \cite{melikhov1, melikhov2, melikhov3, wirbel, jaus2}, QCD sum rules \cite{ball1, ball2, ball3, colangelo, kiselev}, lattice QCD calculations \cite{flynn, abada, ukqcd}, chiral perturbation theory \cite{casal, du}, and the light-front quark model (LFQM) \cite{choi2, choi3, choi4, choi5, cheng1, geng2, cheung, geng3, geng4} have been applied to the calculations of hadronic form factors for rare semileptonic $B$ decays. Experimentally, a significant effort has been made for the advancement of our knowledge of the flavor structure of the SM through the studies of inclusive \cite{alam} as well as exclusive \cite{amar} rare $B$ decays. The violation of CP symmetry in $B$ meson decays was first observed in 2001 (other than in neutral $K$ meson decays) by two experiments: the Belle experiment at KEK and the Babar experiment at SLAC \cite{bevan1}. Both these experiments were constructed and operated on similar time scales and were able to take flavor physics into a new realm of discovery \cite{bevan2}. The Babar and Belle experiments completed taking data in 2008 and 2010 respectively. Recently, numerous measurements of $B$ decays have been performed by the LHC experiments at CERN, in particular, the dedicated $B$ physics experiment LHCb makes a valuable contribution in the understanding of CP violation through the precise determination of the flavor parameters of the SM \cite{lang, adeva, he}.
\par In particular, there has been an enormous interest in studying the decay properties of the $B_c$ meson due to its outstanding properties \cite{gershtein}. Unlike the symmetric heavy quark bound states $b \bar{b}$ (bottomonium) and $c \bar{c}$ (charmonium), $B_c$ meson is the lowest bound state of two heavy quarks ($b$ and $c$) with different flavors and charge. Due to the explicit flavor numbers, $B_c$ mesons can decay only through weak interaction and are stable against strong and electromagnetic interactions, thereby, providing us an opportunity to test the unitarity of CKM quark mixing matrix. The study of an exclusive semileptonic rare $B_c^+ \to D^+ \nu \bar{\nu}$ decay is prominent among all the $B_c$ meson decay modes as it plays a significant role for precision tests of the flavor sector in the SM and its possible NP extensions. At quark level, the decay $B_c^+ \to D^+ \nu \bar{\nu}$ proceeds via $b \to d$ FCNC transition with the intermediate $u$, $c$ and $t$ quarks and most of the contribution comes from the intermediate $t$ quark. Also, due to the neutral and massless final states ($\nu \bar{\nu}$), it opens an unique opportunity to study the $Z$ penguin effects \cite{wick}. As a theoretical input, hadronic matrix elements of quark currents will be required to calculate the transition form factors \cite{cheng2} in order to study the decay rates and branching ratios of the above mentioned decay.
\par The semileptonic rare $B_c^+ \to D^+ \nu \bar{\nu}$ decay has been studied by various theoretical approaches such as constituent quark model (CQM) \cite{geng1}, and QCD sum rules \cite{azizi}. In this work, we choose the framework of light-cone quark model (LCQM) \cite{choi6} for the analysis of this decay process. LCQM deals with the wave function defined on the four-dimensional space-time plane given by the equation $x^{+} = x^{0} + x^{3}$ and includes the important relativistic effects that are neglected in the traditional CQM \cite{brodsky1, lepage}. The kinematic subgroup of the light-cone formalism has the maximum number of interaction free generators in comparison with the point form and instant form \cite{dirac}. The most phenomenal feature of this formalism is the apparent simplicity of the light-cone vacum, because the vacum state of the free Hamiltonian is an exact eigen state of the total light-cone Hamiltonian \cite{brodsky2}. The light-cone Fock space expansion constructed on this vacuum state provides a complete relativistic many-particle basis for a hadron \cite{brodsky3}. The light-cone wave functions providing a description about the hadron in terms of their fundamental quark and gluon degrees of freedom are independent of the hadron momentum making them explicitly Lorentz invariant \cite{brodsky4}.
\par The paper is organized as follows. In Sec. II, we discuss the formalism of light-cone framework and calculate the transition form factors for $B_c^+ \to D^+ \nu \bar{\nu}$ decay process in $q^+ = 0$ frame. In Sec. III, we present our numerical results for the form factors and branching ratios and compare them with other theoretical results. Finally, we conclude in Sec. IV.

\section{Light-cone framework}
In the light-cone framework, we can write the bound state of a meson $M$ consisting of a quark $q_1$ and an antiquark $\bar{q}$ with total momentum $P$ and spin $S$ as \cite{cai}
\begin{eqnarray}
\label{eqn:1}
|M(P,S,S_z)\rangle &=& \int\frac{dp_{q_1}^+d^2\textbf{p}_{q_{1\bot}}}{16\pi^3} \frac{dp_{\bar q}^+d^2\textbf{p}_{\bar q_\bot}}{16\pi^3}16\pi^3 \delta^3(\tilde P-\tilde p_{q_1}-\tilde p_{\bar q}) \nonumber\\ && \times \sum\limits_{\lambda_{q_1},\lambda_{\bar q}}\Psi^{SS_z}(\tilde p_{q_1},\tilde p_{\bar q},\lambda_{q_1},\lambda_{\bar q}) \
 |q_1(p_{q_1},\lambda_{q_1})\bar q(p_{\bar q},\lambda_{\bar q})\rangle,
\end{eqnarray}
where $p_{q_1}$ and $p_{\bar q}$ denote the on-mass shell light-front momenta of the constituent quarks. The four-momentum $\tilde p$ is defined as
\begin{eqnarray}
\label{eqn:2}
\tilde p=(p^+,~\textbf{p}_\perp), \ \textbf{p}_\perp=(p^1,~p^2), \ p^-=\frac{m^2+\textbf{p}_\perp^2}{p^+},
\end{eqnarray}
and
\begin{flalign}
\label{eqn:3}
         |q_1(p_{q_1},\lambda_{q_1})\bar q(p_{\bar q},\lambda_{\bar q})\rangle
        &= b^\dagger(p_{q_1}, \lambda_{q_1})d^\dagger(p_{\bar q}, \lambda_{\bar q})|0\rangle,\nonumber \\
        \{b(p', \lambda'),b^\dagger(p, \lambda)\} &=
        \{d(p', \lambda'),d^\dagger(p, \lambda)\} =
        2(2\pi)^3~\delta^3(\tilde p'-\tilde p)~\delta_{\lambda'\lambda}.
\end{flalign}
The momenta $p_{q_1}$ and $p_{\bar q}$ in terms of light-cone variables are
\begin{eqnarray}
\label{eqn:4}
p_{q_1}^+&=&x_1 P^+, \ \ p_{\bar q}^+=x_2 P^+,\nonumber \\
\textbf{p}_{q_{1\perp}}&=&x_1\textbf{P}_{\perp}+\textbf{k}_{\perp}, \ \ \textbf{p}_{\bar{q}_\perp}=x_2\textbf{P}_{\perp}-\textbf{k}_{\perp},
\end{eqnarray}
where $x_i$ ($i = 1, 2$) represent the light-cone momentum fractions satisfying $x_1 + x_2 = 1$ and $\textbf{k}_{\perp}$ is the relative transverse momentum of the constituent.\\
The momentum-space light-cone wave function $\Psi^{SS_z}$ in Eq. (\ref{eqn:1}) can be expressed as
\begin{eqnarray}
\label{eqn:5}
        \Psi^{SS_z}(\tilde p_{q_1},\tilde p_{\bar q},\lambda_{q_1},\lambda_{\bar q})
                = R^{SS_z}_{\lambda_{q_1}\lambda_{\bar q}}(x,\textbf{k}_\bot)~ \phi(x, \textbf{k}_\bot),
\end{eqnarray}
where $\phi(x, \textbf{k}_\bot)$ describes the momentum distribution of the constituents in the bound state and $R^{SS_z}_{\lambda_{q_1}\lambda_{\bar q}}$ constructs a state of definite spin ($S, S_z$) out of the light-cone helicity ($\lambda_{q_1}, \lambda_{\bar q}$) eigenstates. For convenience, we use the covariant form of $R^{SS_z}_{\lambda_{q_1}\lambda_{\bar q}}$ for pseudoscalar mesons which is given by
\begin{eqnarray}
\label{eqn:6}
        R^{SS_z}_{\lambda_{q_1}\lambda_{\bar q}}(x,\textbf{k}_\bot)
                ={\sqrt{p_{q_1}^+p_{\bar q}^+}\over \sqrt{2} ~{\sqrt{{M_0^2} - (m_{q_1} - m_{\bar q})^2}}}
        ~\bar u(p_{q_1},\lambda_{q_1})\gamma_5 v(p_{\bar q},\lambda_{\bar q}),
\end{eqnarray}
where
\begin{eqnarray}
\label{eqn:7}
M^2_0&=& \frac{m^2_{q_1}+\vec{k}^2_\perp}{x_1}
         + \frac{m^2_{\bar q}+\vec{k}^2_\perp}{x_2}.
\end{eqnarray}
The meson state can be normalized as
\begin{eqnarray}
\label{eqn:8}
        \langle M(P',S',S'_z)|M(P,S,S_z)\rangle = 2(2\pi)^3 P^+
        \delta^3(\tilde P'- \tilde P)\delta_{S'S}\delta_{S'_zS_z}~,
\end{eqnarray}
so that
\begin{eqnarray}
\label{eqn:9}
        \int {dx\,d^2\textbf{k}_\bot\over 2(2\pi)^3}~|\phi(x,\textbf{k}_\bot)|^2 = 1.
\end{eqnarray}
We choose the Gaussian-type wave function to describe the radial wave function $\phi$:
\begin{eqnarray}
\label{eqn:10}
\phi(x,\textbf{k}_\bot)=
\sqrt{\frac{1}{\pi^{3/2}\beta^{3}}}
\exp(-\textbf{k}^{2}/2\beta^{2}),
\end{eqnarray}
where $\beta$ is a scale parameter and $\textbf{k}^2=\textbf{k}^2_\bot + k^2_z$ denotes the internal momentum of meson. The longitudinal component $k_z$ is defined as
\begin{eqnarray}
\label{eqn:11}
k_z=(x-\frac{1}{2})M_0 + \frac{m^2_{q_1}-m^2_{\bar q}}{2M_0}.
\end{eqnarray}

\subsection{Form factors for the semileptonic $B_c^+ \to D^+ \nu \bar{\nu}$ decay in LCQM}
The form factor $f_+ (q^2)$ and $f_T (q^2)$ can be obtained in $q^+ = 0$ frame with the ``good" component of current, i.e. $\mu = +$, from the hadronic matrix elements given by \cite{cai}
\begin{eqnarray}
\label{eqn:12}
\langle D^+|\bar d \gamma^\mu b |B_c^+\rangle = f_+(q^2) P^\mu + f_-(q^2) q^\mu,
\end{eqnarray}
and
\begin{eqnarray}
\label{eqn:13}
\langle D^+|\bar{d}i\sigma^{\mu\nu}q_\nu b|B_c^+\rangle
= \frac{f_T(q^2)}{(M_{B_c^+}+M_{D^+})}[q^2 P^\mu - (M^2_{B_c^+}-M^2_{D^+})q^\mu].
\end{eqnarray}
It is more convenient to express the matrix element defined by Eq. (\ref{eqn:12}) in terms of $f_+(q^2)$ and $f_0(q^2)$ as:
\begin{eqnarray}
\label{eqn:112}
\langle D^+|\bar d \gamma^\mu b |B_c^+\rangle = F_+(q^2)\left[P^\mu
-\frac{M^2_{B_c^+}-M^2_{D^+}} {q^2} q^\mu\right] + f_0(q^2)\frac{M^2_{B_c^+}-M^2_{D^+}}{q^2}q^\mu,
\end{eqnarray}
with
\begin{equation}
F_+ (q^2) = f_+(q^2) \quad \textrm{and} \quad
f_0 (q^2) = f_+ (q^2) + \frac{q^2}{M^2_{B_c^+}-M^2_{D^+}} f_- (q^2).\nonumber
\end{equation}
Here $P = P_{B_c^+} + P_{D^+}$ and $q = P_{B_c^+} - P_{D^+}$ and $0\leq q^2\leq (M_{B_c^+}-M_{D^+})^2$. \\
Using the parameters of $b$ and  $d$ quarks, the form factors $f_+(q^2)$ and $f_T(q^2)$ can be respectively expressed in the quark explicit forms as follows \cite{choi6}
\begin{eqnarray}
\label{eqn:14}
f_{+}(q^{2}) =
\int^{1}_{0}dx\int d^{2}{\textbf{k}_\bot}
\sqrt{\frac{\partial k'_z}{\partial x}}
\sqrt{\frac{\partial k_z}{\partial x}} \phi_{d}(x,{\textbf{k}'_\bot}) \phi_{b}(x,{\textbf{k}_\bot})
\frac{A_{b} A_{d}+{\textbf{k}_\bot}\cdot{\textbf{k}'_\bot}}
{ \sqrt{ A_{b}^{2}+ \textbf{k}_\bot^{2}}\sqrt{ A_{d}^{2}+
\textbf{k}_\bot^{'2}} },
\end{eqnarray}
and
\begin{eqnarray}
\label{eqn:15}
f_{T}(q^{2}) &=& -\int^{1}_{0}dx\int d^{2}{\textbf{k}_\bot}
\sqrt{\frac{\partial k'_z}{\partial x}}
\sqrt{\frac{\partial k_z}{\partial x}} \phi_{d}(x,{\textbf{k}'_\bot}) \phi_{b}(x,{\textbf{k}_\bot}) \nonumber\\
&& \times \frac{x (M_{B_c^+}+M_{D^+}) \biggl[(m_d-m_b)\frac{\textbf{k}_\bot\cdot\textbf{q}_\bot}{\textbf{q}^2_\perp}
+ A_b\biggr]}
{ \sqrt{ A_{b}^{2}+ \textbf{k}_\bot^{2}}\sqrt{ A_{d}^{2}+
\textbf{k}_\bot^{'2}} },
\end{eqnarray}
where ${\textbf{k}'_\bot}={\textbf{k}_\bot}-x{\textbf{q}_\bot}$ represents the final state transverse momentum, $A_{b}=xm_{b} + (1-x)m_{\bar{q}}$ and $A_{d}=xm_{d} + (1-x)m_{\bar{q}}$. The term ${\partial k_{z}}/{\partial x}$ denotes the Jacobian of
the variable transformation $\{x,{\textbf{k}_\bot}\}\to {\textbf{k}}=
(k_{z},{\textbf{k}_\bot})$. \\
The LCQM calculations of form factors have been performed in the $q^+ = 0$ frame \cite{drell, west}, where $q^2 = q^+ q^- - \textbf{q}^2_\bot = -\textbf{q}^2_\bot < 0$ (spacelike region). The calculations are analytically continued to the $q^2 >0$ (timelike) region by replacing $\textbf{q}_\bot$ to  $i\textbf{q}_\bot$ in the form factors. To obtain the numerical results of the form factors, we use the change of variables as follows
\begin{eqnarray}
\label{eqn:16}
\textbf{k}_\bot&=& \bm{\ell}_\perp
+\frac{x\beta^2_{B_c^+}}{\beta^2_{B_c^+}+\beta^2_{D^+}}\textbf{q}_\bot,
\nonumber\\
\textbf{k}'_\bot&=&\bm{\ell}_\perp
-\frac{x\beta^2_{D^+}}{\beta^2_{B_c^+}+\beta^2_{D^+}}\textbf{q}_\bot.
\end{eqnarray}
The detailed procedure of analytic solutions for the weak form factors in timelike region has been discussed in literature \cite{choi7}.\\
For the sake of completeness and to compare our analytic solutions, we use a double pole parametric form of form factors expressed as \cite{geng1}:
\begin{eqnarray}
\label{eqn:17}
f(q^2) = \frac{f(0)}{1 + {\cal{A}} \ s + {\cal{B}} \ s^2},
\end{eqnarray}
where $s = q^2/M_{B_c^+}^2$, $f(q^2)$ denotes any of the form factors, $f(0)$ denotes the form factors at $q^2 = 0$. Here ${\cal A}$, ${\cal B}$ are the parameters to be fitted from Eq. (\ref{eqn:17}). While performing calculations, we first compute the values of $f_+(q^2)$ and $f_T(q^2)$ from Eqs. (\ref{eqn:14}) and (\ref{eqn:15}) in $0\leq q^2\leq (M_{B_c^+}-M_{D^+})^2$, followed by extraction of the parameters ${\cal A}$ and ${\cal B}$ using the values of $M_{B_c^+}$ and $f(0)$, and then finally fit the data in terms of parametric form.

\subsection{Decay rate and Branching ratio for $B_c^+ \to D^+ \nu \bar{\nu}$ decay}
\begin{figure}[h]
  \begin{center}
    \includegraphics[width=4 in]{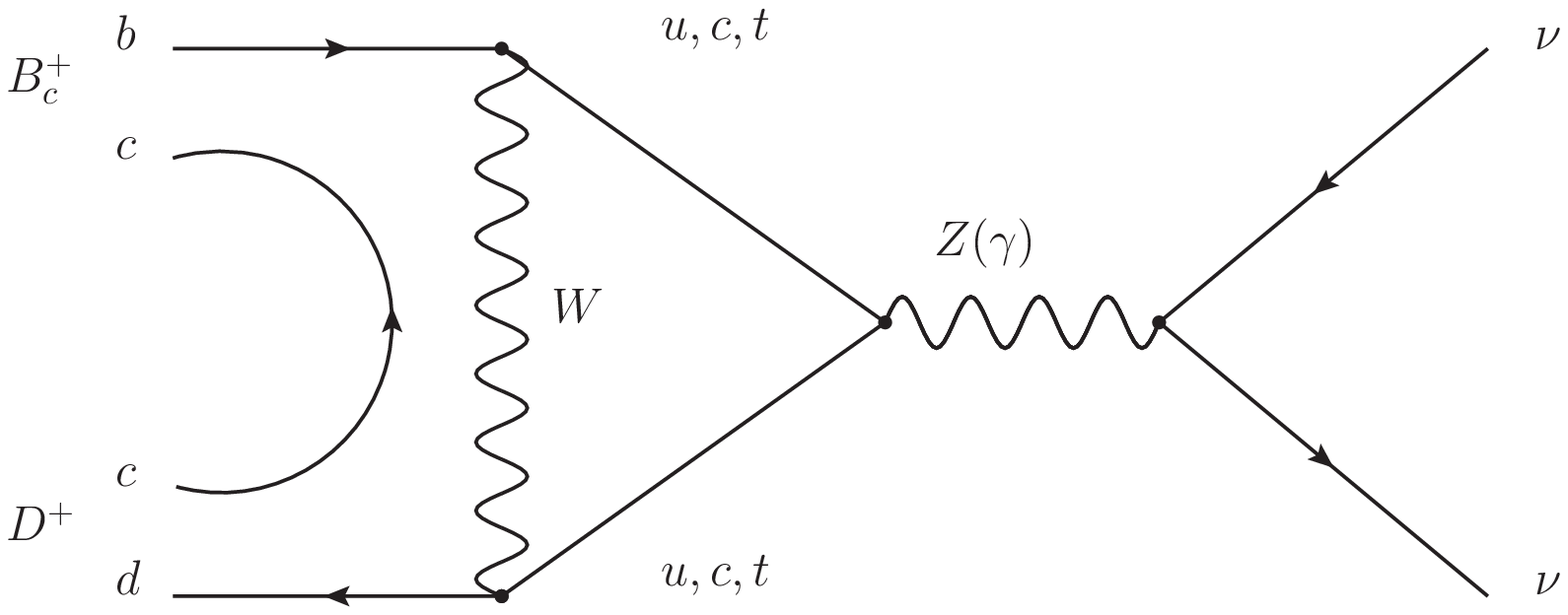}
    \includegraphics[width=4 in]{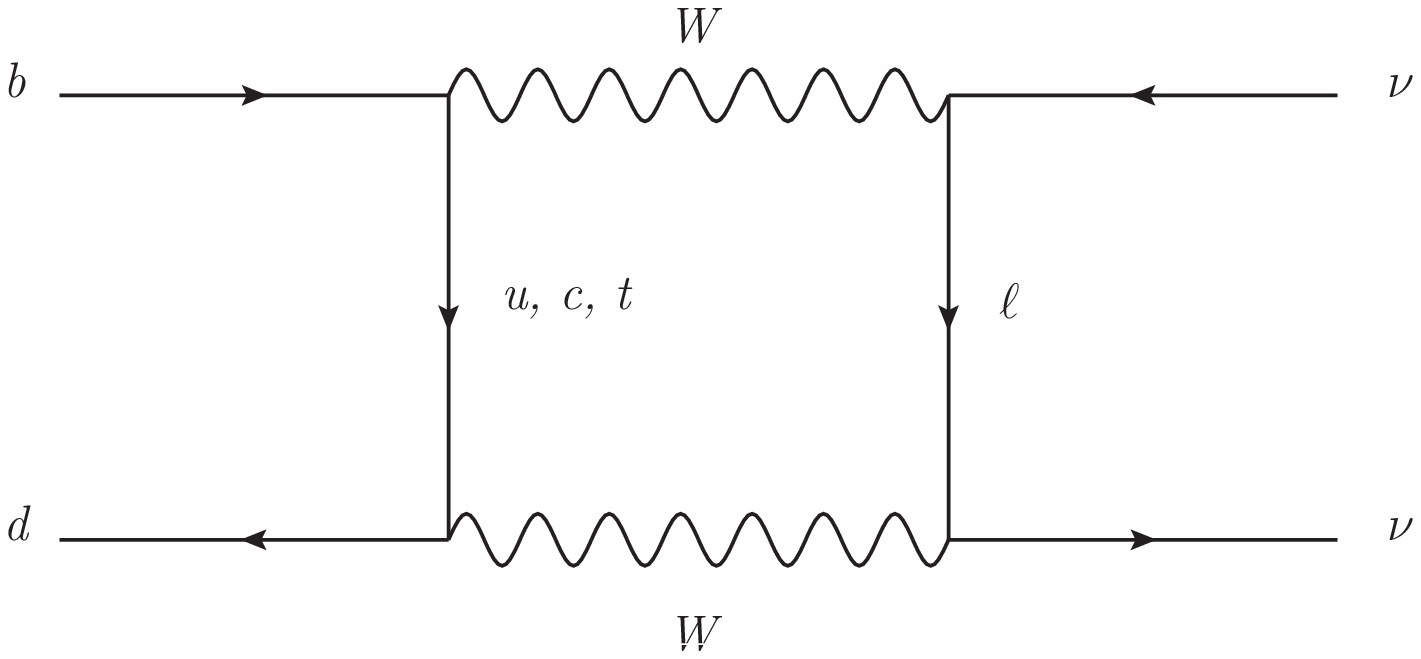}
  \end{center}
  \caption{Loop diagrams for $B_c^+ \to D^+ \nu \bar{\nu}$ decay process.}
  \label{di1}
\end{figure}

At the quark level, the rare semileptonic $B_c^+ \to D^+ \nu \bar{\nu}$ decay is described by the $b \to d$ FCNC transition. As mentioned earlier, these kind of transitions are forbidden at the tree level in the SM and occur only through loop diagrams as shown in the Fig. \ref{di1}. They receive contributions from the penguin and box diagrams \cite{geng1}. Theoretical investigation of these rare transitions usually depends on the effective Hamiltonian density. The effective interacting Hamiltonian density responsible for $b \to d$ transition is given by \cite{grin}:
\begin{eqnarray}
\label{eqn:18}
\mathcal{H}_{\rm eff} (b \to d \nu \bar{\nu})=\frac{G_F}{\sqrt{2}} \frac{\alpha V_{tb}V^*_{td}}{2 \pi \textrm{sin}^2\theta_W} X(x_t) \bar{d} \gamma_{\mu} (1 - \gamma_5) b \bar{\nu} \gamma^{\mu} (1 - \gamma_5) \nu,
\end{eqnarray}
where $G_F$ is the Fermi constant, $\alpha$ is the electromagnetic fine structure constant, $\theta_W$ is the Weinberg angle, $V_{ij}$ ($i=t, j=b \ \textrm{and} \ d$) are the CKM matrix elements and $x_t = m_{t}^2/M_{W}^2$.
\newpage
The function $X (x_t)$ denotes the top quark loop function, which is given by
\begin{eqnarray}
\label{eqn:19}
X (x_t) = \frac{x_t}{8} \biggl(\frac{2 + x_t}{x_t - 1} + \frac{3 x_t - 6}{(x_t - 1)^2} {\rm ln}\,x_t\biggr).
\end{eqnarray}
\par The differential decay rate for $B_c^+ \to D^+ \nu \bar{\nu}$ can be expressed in terms of the form factors as \cite{choi6}
\begin{eqnarray}
\label{eqn:20}
\frac{d\Gamma}{ds} =\frac{M^5_{B_c^+} G^2_F}{2^8\pi^5 \textrm{sin}^4\theta_W}\alpha^2
|V_{tb}V^*_{td}|^2 |X(x_t)|^2\phi^{3/2}_{D^+} |f_+|^2,
\end{eqnarray}
where $\phi _{D^+} = (1 - r_{D^+})^2 - 2 s (1 + r_{D^+}) + s^2$ with $s = q^2/M_{B_c^+}^2$ and $r_{D^+} = M_{D^+}^2/M_{B_c^+}^2$.\\
The differential branching ratio ($d \textrm{BR}/ds$) can be obtained by dividing the differential decay rate ($d \Gamma/ds$) by the total width ($\Gamma_{\textrm{total}}$) of the $B_c^+$ meson and then by integrating the differential branching ratio over $s = q^2/M_{B_c^+}^2$, we can obtain the branching ratio (BR) for $B_c^+ \to D^+ \nu \bar{\nu}$ decay.
\section{Numerical Results}
\begin{figure} [h]
  \begin{center}
    \includegraphics[width=4 in]{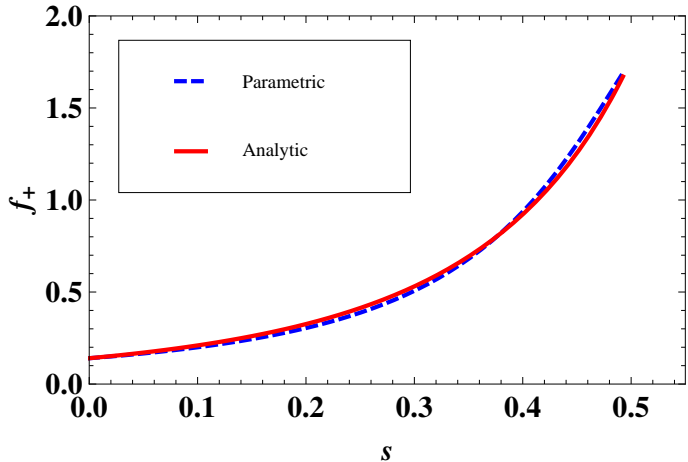}
  \end{center}
  \caption{Analytic solutions of $f_+$ (thick solid curve) compared with the parametric results (dashed curve), with defination $s = q^2/M_{B_c^+}^2$.}
  \label{fig:1}
\end{figure}

\begin{figure} [h]
  \begin{center}
    \includegraphics[width=4.2 in]{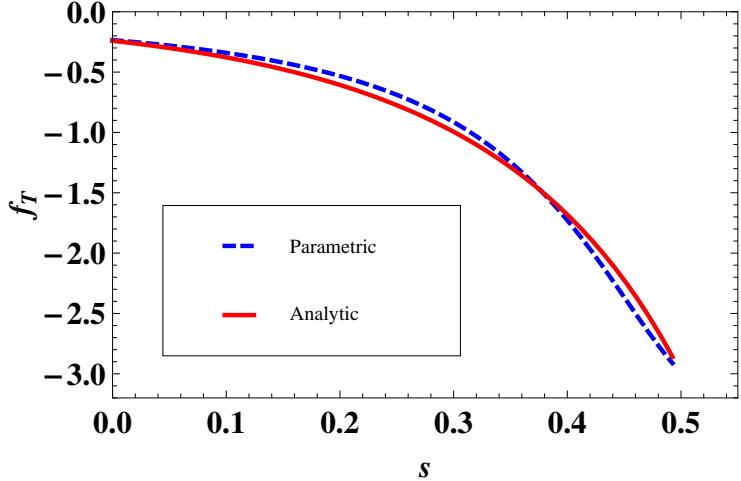}
  \end{center}
  \caption{Analytic solutions of $f_T$ (thick solid curve) compared with the parametric results (dashed curve), with defination $s = q^2/M_{B_c^+}^2$.}
  \label{fig:2}
\end{figure}

Before obtaining the numerical results of the form factors for the semileptonic $B_c^+ \to D^+ \nu \bar{\nu}$ decay, we first specify the parameters appearing in the wave functions of the hadrons. We have used the constituent quark masses as \cite{wang, cai}
\begin{center}
$m_b = 4.8$ GeV, $m_d = 0.25$ GeV and $m_c = 1.4$ GeV.
\end{center}
The parameter $\beta$ that describes the momenta distribution of constituent quarks can be fixed by the meson decay constants $f_{B_c^+}$ and $f_{D^+}$, respectively. The $\beta$ parameters that we have used in our work are given as \cite{geng1}
\begin{center}
$\beta_{B_c^+} = 0.81$ GeV and $\beta_{D^+} = 0.46$ GeV.
\end{center}
Using the above parameters, we present the analytic solutions of the form factors $f_+$ and $f_T$ (thick solid curve) for $0 \leq q^2 \leq (M_{B_c^+} - M_{D^+})^2$ in Figs. \ref{fig:1} and \ref{fig:2}, respectively. We have also shown the results obtained from the parametric formula (dashed curve) given by Eq. (\ref{eqn:17}).
We would like to mention here that the point $q^2 = 0$ represents the maximum recoil point and the point $q^2 = q^2_{\textrm{max.}} = (M_{B_c^+} - M_{D^+})^2$ represents the zero recoil point where the produced meson is at rest.
As we can see from Figs. \ref{fig:1} and \ref{fig:2}, the form factors $f_+$ and $f_T$ increases and decreases exponentially with respect to $q^2$. The analytic solutions of form factors given by Eqs. (\ref{eqn:14}) and (\ref{eqn:15}) are well approximated by the parametric form in the physical decay region $0 \leq q^2 \leq (M_{B_c^+} - M_{D^+})^2$. For a deeper understanding of the results, we have listed the numerical results for the form factors $f_+$ and $f_T$ at $q^2 = 0$ and the parameters ${\cal A}$ and ${\cal B}$ of the double pole form in Table \ref{tab1}. For the sake of comparison, we have also presented the results of other theoretical models.
\begin{table} [h]
\centering
\caption{Form factors for $B_c^+ \to D^+ \nu \bar{\nu}$ decay process at $q^2 = 0$ and the parameters ${\cal A}$ and ${\cal B}$ defined by Eq. (\ref{eqn:17}) and their comparison with other theoretical model predictions.}
\begin{tabular*}{\textwidth}{@{}l*{15}{@{\extracolsep{0pt plus
12pt}}r}}
\hline\hline\\[-1.5 ex]
Model & $f_+ (0)$ & ${\cal A}$ & ${\cal B}$ & $f_T (0)$ & ${\cal A}$ & ${\cal B}$\\[1 ex]
\hline\\[-1.5 ex]
This work & 0.140 & $-3.263$ & 2.846 & $-0.234$ & $-3.430$ & 3.174\\[1 ex]
CQM \cite{geng1} & 0.123 & $-3.35$ & 3.03 & $-0.186$ & $-3.52$ & 3.38\\[1 ex]
SR \cite{azizi} & 0.22 & $-1.10$ & $-2.48$ & $-0.27$ & $-0.72$ & $-3.24$\\[1 ex]
Linear \cite{choi6} & 0.086 & $-3.50$ & 3.30 & $-0.120$ & $-3.35$ & 3.06\\[1 ex]
HO \cite{choi6} & 0.079 & $-3.20$ & 2.81 & $-0.108$ & $-3.18$ & 2.77\\[1 ex]
\hline\\[-3.28 ex]\hline
\end{tabular*}
\label{tab1}
\end{table}
It can be seen from the Table that the values of form factors $f_+$ and $f_T$ at $q^2 = 0$ in our model agree quite well with the CQM. The difference in the values with respect to other models might be due to the different assumptions of the models or  different choices of parameters.
\par
\begin{figure} [h]
  \begin{center}
    \includegraphics[width=4 in]{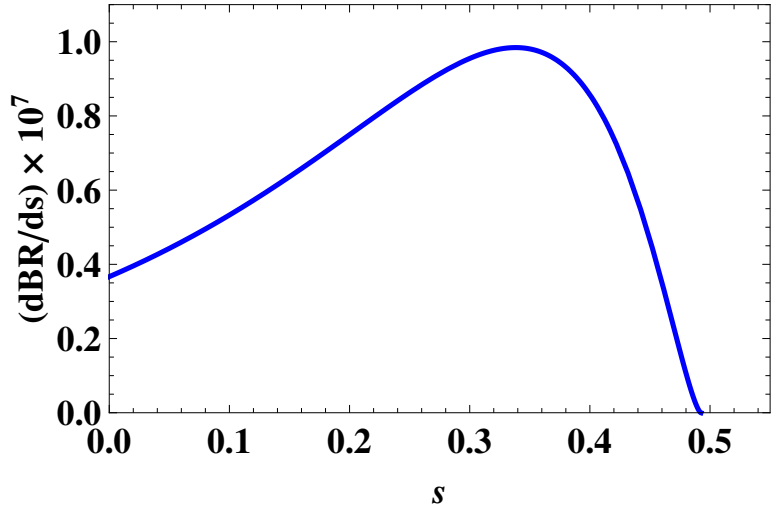}
  \end{center}
  \caption{Differential branching ratios as a function of $s$ for $B_c^+ \to D^+ \nu \bar{\nu}$ decay.}
  \label{fig:3}
\end{figure}

 To estimate the numerical value of the branching ratio for $B_c^+ \to D^+ \nu \bar{\nu}$ decay (defined in Eq. (\ref{eqn:20})), the various input parameters used are \cite{choi6} $\alpha^{-1} = 129$, $|V_{tb}V^*_{td}| = 0.008$, $M_W = 80.43$ GeV, $m_t = 171.3$ GeV and $\textrm{sin}^2\theta_{W} = 0.2233$. The lifetime of $B_c^+$ ($\tau_{B_c^+} = 0.507$ ps) is taken from the Particle Data Group \cite{patri}. Our results for the differential branching ratio as a function of $s$ is shown in Fig. \ref{fig:3}.
\par Our prediction for the decay branching ratio of $B_c^+ \to D^+ \nu \bar{\nu}$ decay is listed in Table \ref{tab2} and compared with the other theoretical predictions.
\begin{table} [h]
\centering
\caption{Branching ratio for $B_c^+ \to D^+ \nu \bar{\nu}$ decay in LCQM and its comparison with the other models }
\begin{tabular}{p{2in}c}
\hline\hline\\[-1.5 ex]
Model & Branching ratios (in units of $10^{-8}$)\\[1 ex]
\hline\\[-1.5 ex]
This work & 3.33\\[1 ex]
CQM \cite{geng1} & 2.74\\[1 ex]
QCD sum rules \cite{azizi} & 3.38\\[1 ex]
Linear \cite{choi6} & 1.31\\[1 ex]
HO \cite{choi6} & 0.81\\[1 ex]
\hline\\[-3.55 ex]\hline
\end{tabular}
\label{tab2}
\end{table}
As we can see from Table \ref{tab2}, the result predicted by LCQM approximately agrees with the prediction given by QCD sum rules whereas it is slightly larger when compared with the results of CQM. At present, we do not have any deep understanding of these
values, however they do indicate that these results may be important even in a more rigorous model. The measurements can perhaps be substantiated by measurement of the decay width of $B$ mesons. Several experiments at LHCb are contemplating the possibility of searching for more $B$ meson decays.

\section{Conclusions}
We have studied the exclusive semileptonic rare $B_c^+ \to D^+ \nu \bar{\nu}$ decay within the framework of LCQM. In our analysis, we have evaluated the transition form factors $f_+ (q^2)$ and $f_T (q^2)$ in the $q^{+} = 0$ frame and then extended them from the spacelike region ($q^2 < 0$) to the timelike region ($q^2 >0$) through the method of analytical continuation using the constituent quark masses ($m_b$, $m_d$ and $m_c$) and the parameters describing the momentum distribution of the constituent quarks ($\beta_{B_c^+}$ and $\beta_{D^+}$), respectively. The numerical values of $\beta_{B_c^+}$ and $\beta_{D^+}$ have been fixed from the meson decay constants $f_{B_c^+}$ and $f_{D^+}$, respectively. We have also compared the analytic solutions of transition form factors with the results obtained for the form factors using the double pole parametric form. Using the numerical results of transition form factors, we have calculated the decay branching ratio and compared our result with the other theoretical model predictions. The LCQM result for the decay branching ratio of $B_c^+ \to D^+ \nu \bar{\nu}$ decay comes out to be $3.33 \times 10^{-8}$ which approximately agrees with the prediction given by QCD sum rules \cite{azizi}. This result can also be tested at the LHCb experiments in near future. 

To conclude,  new experiments aimed at measuring the decay branching ratios are not only needed for the profound understanding of $B$ decays but also to restrict the model parameters for getting better knowledge on testing the unitarity of CKM quark mixing matrix. This will provide us an useful insight into the phenomenon of CP violation.

\section*{Acknowledgements}
Authors would like to acknowledge Chueng-Ryong Ji (North Carolina State University, Raleigh, NC) for the helpful discussions and Department of Science and Technology (Ref No. SB/S2/HEP-004/2013) Government of India for financial support.

\end{document}